\documentclass[twoside,a4paper,11pt]{sca}
\usepackage{graphicx}
\usepackage{hyperref}
\usepackage{natbib}  
\begin{document}
\pagenumbering{arabic}
\pagestyle{myheadings}
\thispagestyle{empty}
{\flushright\includegraphics[width=\textwidth,bb=90 650 520 700]{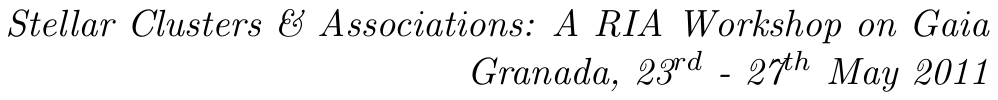}}
\vspace*{0.2cm}
\begin{flushleft}
{\bf {\LARGE
%
Star cluster formation and some implications for GAIA
%
}\\
\vspace*{1cm}
%
Pavel Kroupa
%
}\\
\vspace*{0.5cm}
%
University of Bonn, Auf dem H\"ugel 71, 53121 Bonn, Germany
%
\end{flushleft}
%
\markboth{
Some implications of star clusters for GAIA
}{ 
%
Pavel Kroupa
%
}
\thispagestyle{empty}
\vspace*{0.4cm}
\begin{minipage}[l]{0.09\textwidth}
\ 
\end{minipage}
\begin{minipage}[r]{0.9\textwidth}
\vspace{1cm}
\section*{Abstract}{\small
Stars form in spatially and temporarily correlated star formation
events (CSFEs) and the dynamical processes within these ``embedded
clusters'' leave imprints in the stellar populations in galactic
fields. Such imprints are correlations in phase space
(e.g. gravitationally bound star clusters, tidal streams), in the
binary properties of stars and in the present-day stellar mass
functions in the surviving clusters. The dynamical processes include
expulsion of massive stars from cluster cores, disruption of CSFEs due
to residual gas expulsion and energy-equipartition driven evaporation
of stars from clusters leading to dark star clusters and cold kinematical
streams with epicyclic overdensities. The properties of such
phase-space structures in the Milky Way (MW) field depend on the
effective gravitational potential of the MW.  GAIA data will
significantly constrain all of these aspects, and will in particular
impact on gravitational dynamics via the properties of cold streams
and on star-formation via the constraint on the gas expulsion process
through the expanding unbound populations that must be associated with
every CSFE.
%
\normalsize}
\end{minipage}
%
%
%
\section{Introduction \label{intro}}

The data generated by the GAIA mission will provide historically
unprecedented constraints on the formation and evolution of the MW and
will also constrain gravitational theories in the ultra-weak field
limit. In order to interpret the GAIA data we will require detailed
knowledge on how the stellar phase-space distribution function, which
defines the properties of the MW, comes about. This is a complex
problem: To interpret the data we will need to compute models of how
stars populate phase space. But in which dynamical theory? If we
assume Newtonian dynamics the computations can be done today but are
equivalent to assuming the concordance cosmological model to be the
correct description of reality. The current astronomical constraints
already suggest it to be excluded though \citep{PN10, Kroupa10}, and
the so computed model MW would then not match the GAIA constraints
unless some additional unknown physics is invoked. Alternatively,
perhaps reality follows Modified Newtonian Dynamics (MOND,
\citealt{Milgrom09}) or another alternative (e.g. \citealt{Moffat09})
in which case it may be possible to naturally obtain consistency of
the data with the model. But, at present it is not possible to compute
a dynamical model of the MW from scratch (i.e. the stars spreading
into the field from their birth structures) with Milgrom's
dynamics. Until this becomes possible, we will not be able to
correctly account for the evolution of the MW, since we will always
end up with the trivial solution that the MW is described by Newton's
laws plus ad hoc dark unknowns, in order to match the GAIA data.

Irrespective of these issues, here an outline is provided of how the
signatures of CSFEs are likely to be contained in astrometric and
stellar-population survey data such as will become available with the
GAIA mission.

\section{Do all stars from in star clusters?}

The phase space distribution of stars in the MW is not only defined by
the effective potential of the MW but also by the dynamical origin of
the stars. Here a fundamental {\sc Question} arises: {\it Do all stars
  form in embedded star clusters} \citep{Bressert10}? If this were the
case then the stellar dynamical processes within the clusters need to
be treated first in order to understand the properties of the
phase-space distribution function of stars in the MW field.  Each
cluster of a given energy scale (given by its mass and radius)
provides a unique contribution to the stellar population in the field.
The concepts of {\it kinematical population synthesis}
\citep{Kroupa02}, {\it dynamical population synthesis}
\citep{K95d,MK11} and of the {\it integrated galactic initial mass
  function} (IGIMF) of stars \citep{KW03,WKP11} arise from the concept
of adding up the contributions of all clusters.

Rather than referring to embedded star clusters, the term {\it
  Correlated Star Formation Event} (CSFE) is applied here.  A CSFE
means a group (or ``cluster'') of stars formed over a spatial scale of
about one~pc within about one~Myr. Observations suggest that these are
typical star-forming structures, ranging from the Taurus-Auriga
sub-clusters of a dozen binaries in each, through dense-populous young
clusters such as the Orion Nebula cluster (ONC), to massive
star-bursting embedded clusters with masses $>10^6\,M_\odot$. With the
work of \cite{Larsen04} it has become clear that there is no special
mode of globular cluster (GC) formation, but that there is as
continuous distribution of star-forming events from the smallest to
the most massive CSFEs.  A CSFE does not have to be gravitationally
bound in entirety.

Even if a CSFE were to be formally a gravitationally self bound
structure (i.e. a classical star cluster), then a certain fraction of
its stars will always be below a density threshold in its outer
regions. An observer who applies a density threshold to define
``clustered star formation'', implying that stars found in regions
below this threshold would be termed stars formed in isolation, would
thus always find isolated stars in star-forming molecular clouds.  For
example, an individual cluster represented by a Plummer model in
virial equilibrium, i.e. being gravitationally bound, with a Plummer
radius of~0.3~pc and with a mass of $20, 100, 1000\,M_\odot$ has,
respectively, 70, 30, 10~per cent of its stars outside a radius at
which the projected surface density is 60~stars/pc$^2$
(Pflamm-Altenburg, priv.comm.). The ``efficiency'' of clustered star
formation thus appears to increase with CSFE mass.  A whole population
of 0.3~pc~Plummer CSFEs ranging from $10\,M_\odot$ to $10^4\,M_\odot$
has a fraction of 30~per cent ``isolated'' stars. Exponential density
profiles (rather than Plummer profiles) or expanding stellar
populations after the expulsion of residual gas from initially bound
CSFEs would change these numbers such that the fraction may be larger
or smaller depending on the characteristic length scale of the CSFEs,
such that the observed fraction of star-formation in dense clusters
\citep{Bressert10} may be arrived at. The fraction of stars which are
apparently formed outside of dense clusters is enhanced due to
low-mass embedded clusters dispersing on a few crossing time scales
and stars being ejected from dense regions of their CSFEs
\citep{WBM11}.  Despite the significant fraction of ``isolated
stars'', the stellar-dynamics of such CSFEs remains that of an
embedded star cluster.

Thus the above {\sc Question} must be rephrased to the following form:
{\it Do all stars from in CSFEs?}. The answer is yes, because CSFEs
are confined to the dense cores of molecular clouds which is where
stars form. The least-massive CSFEs can be termed to be non-clustered
star formation if these include groups of a few low-mass stars. With
this answer, GAIA-relevant modelling can be performed on galaxy scales
because the quantities of interest (e.g. velocity and space
distributions, binary populations, stellar mass functions) are then
given by time-dependent integrals over all CSFEs, once we know the
mass and size distribution of the CSFEs.

This contribution touches upon the issue of how star formation in
CSFEs affects various aspects of galactic astrophysics. A fundamental
concept underlying the approach is that to make a galaxy one merely
needs to add up all CSFEs over time (the {\it cosmological lego},
i.e. the concept that embedded star-clusters are the fundamental
building blocks of galaxies \citealt{Kroupa05}).

\section{Formation of CSFEs}

It is by now well established observationally that a dense molecular
cloud region of a typical scale of a few~pc fragments into gas rich
sub-clusters with individual diameters of less than a~pc
\citep{Lada_Lada03}.  Numerical simulations of turbulent
self-gravitating clouds also show this process
(e.g. \citealt{Klessen04}). This phase of star-cluster formation takes
about 1~Myr by which time between about~10 and~30~per cent of the gas
has formed stars with total mass $M_{\rm ecl}$ distributed according
to the stellar IMF with a most-massive-star---star-cluster mass
relation as a function of the time $t$,
\citep{BVB04,Peters10,Peters11a}
\begin{equation}
m_{\rm max}(t) = 0.39\,M_{\rm ecl}(t)^{2/3},
\label{eq:mmax_pred}
\end{equation}
in good agreement with the observed relation, $m_{\rm max}={\rm
  fn}(M_{\rm ecl})$, established after star formation has ended
\citep{WKB10}.

Once the stellar feedback is sufficient to significantly impact the
molecular cloud core the gas is blown out. An issue not fully answered
yet is whether gas blow-out is explosive (i.e. faster than a cluster
dynamical time) or whether it is adiabatic, and how this depends on
cluster mass (see below). 

Nevertheless, the theoretical problem of star formation can be divided
into two computationally accessible phases.  In the first phase,
gravo-hydrodynamical simulations can be used to study the collapse and
fragmentation of the cloud, while in the second phase purely
stellar-dynamical ($N$-body) methods need to be used to treat the
stellar-dynamical processes with high accuracy. The transition between
the gas-dominated and the pure $N$-body dynamics dominated phases
occurs at an age of about one~Myr and can only be treated with
simplifying assumptions, especially if substantial CSFEs with $M_{\rm
  ecl}>10^3\,M_\odot$ are to be studied. Here the sole computationally
feasible approach is to model the time-varying gas potential as an
additive analytical back ground potential
(e.g. \citealt{GeyerBurkert01, KAH}) allowing major scans of parameter
space \citep{BK07}. Since the formation of each individual star takes
about~0.1~Myr, most stars have been born by~1~Myr. For CSFEs with
$M_{\rm ecl}>\,{\rm few}\,100\,M_\odot$ they have had enough time to
orbit through the CSFE once at least \citep{Kroupa05} such that it can
be assumed that such CSFEs are close to dynamical equilibrium when the
gas is blown out.

\section{Early stellar dynamical processes}

An initially sub-structured stellar population within a CSFE can merge
to an embedded cluster on a crossing time scale, or it can disperse
into the field, depending on the velocity field of the cloud and the
stellar feedback processes \citep{McMillan07, Clark08, Fellhauer09,
  Bonnell11}. The individual sub-clusters are likely to be initially
mass segregated because the most massive stars form in the densest
regions \citep{Maschberger10}. Even if the overall structure is not
mass segregated, the mass segregation time scale is short and of the
order of
\begin{equation}
t_{\rm ms}={\mathcal O}({m_{\rm av} \over m_{\rm mass}}\,t_{\rm relax}),
\end{equation}
where $m_{\rm av}, m_{\rm mass}$ are the average stellar mass and the
mass of the massive star, respectively, while $t_{\rm relax}$ is the
two-body relaxation time. For example, for the ONC $t_{\rm
  relax}\approx 0.6\,$Myr such that $t_{\rm ms}\approx0.12\,$Myr$\ll
\,$age of the ONC \citep{Kroupa08}.

Once a few massive stars collect in the cluster core the core decays
by ejecting the stars on a time scale 
\begin{equation}
t_{\rm decay}={\mathcal O}\left( N_{\rm m}\,t_{\rm core, cross} \right)
\end{equation}
where $N_{\rm m}$ is the number of massive stars in the core and
$t_{\rm core, cross}$ is the core-crossing time. For the ONC the core
has a radius of about~0.02~pc and the mass of the core is about
$150\,M_\odot$ such that $t_{\rm core, cross}\approx 10^4\,$yr and
$t_{\rm decay}\approx 10^5\,$yr$\;\ll$ age of the ONC. Thus, dense
clusters are efficient accelerators of massive stars such that a large
fraction of them get dispersed into the galactic field as soon a core
of massive stars forms \citep{PAK06, BKO11,FPZ11,GG11}. Since most
massive stars are born as multiple systems \citep{SE10} the dynamical
encounters within the cores can be violent.

Indeed, initially mass segregated clusters in which all massive stars
are massive binaries eject a large fraction of their massive stars
within 3~Myr, as is demonstrated by the extensive young star-cluster
$N$-body library of mass segregated and not mass segregated clusters
with and without binaries (Oh et al., in prep.).  The primary of an
ejected massive binary ultimately explodes as a core-collapse
supernova releasing the massive companion into a random
direction. About 1-4~per cent of all ejected massive stars can
therefore not be traced back to their cluster of origin \citep{PAK10},
while the observed fraction of massive stars that appear to have
formed in isolation is about~1~per cent only. Therefore there is no
evidence for the formation of massive stars in isolation.

Concerning the binary population in clusters, it is well know from
$N$-body work that binary stars can get disrupted but may also harden
through stellar-dynamical encounters
\citep{K95d,Portgies01,Kaczmarek11}.  It has been shown that a unified
invariant birth binary population in which every star with mass
$<\,$few$\,M_\odot$ is in a binary naturally evolves into the observed
binary population which has a smaller binary fraction
\citep{MKO11}. By summing all initial populations transformed by all
CSFEs the observed period and mass-ratio distributions of the MW field
are accounted for naturally through the stellar-dynamical disruptions
of the birth population within the individual CSFEs \citep{MK11}.

The expansion of the cluster associated with the dynamical activity of
the core is not significant compared to the expansion induced due to
the expulsion of residual gas.

\section{The star formation history within a cluster}

Some low-mass clusters show evidence for a significant age spread of
their stars \citep{Palla07}.  This can be interpreted to have arisen
through a slow formation time-scale of up to 10~Myr and would be in
contradiction with cluster formation on a dynamical time-scale
\citep{Hartmann01}.

There are two processes which naturally lead to an age spread in star
clusters such that CSFE formation can occur on a dynamical time but
still contain older stars:

\begin{enumerate}

\item As the pre-cluster cloud core collapses on a dynamical time scale
the potential deepens and young nearby stars from a surrounding older
association can be captured to become cluster members.  In this way
8~per cent of all ONC stars may be captured older stars \citep{PAK07},
while 30~per cent or more of stars in $\omega$~Cen may be captured
from the originally hosting dwarf galaxy \citep{FPE06}

\item After its massive stars have died a young cluster can re-accrete
  gas if it enters a molecular cloud, because the cluster potential
  leads to a hydrodynamical instability \citep{PAK09}.

\end{enumerate}
For massive clusters (about$\,>10^6\,M_\odot$) the gas cannot be
removed readily due to the deep potential and shock thermalisation of
stellar winds such that successive populations of stars may emerge
\citep{TenarioTagle03,Wuensch08}. How the observed peculiar chemical
compositions in the massive GCs may be accounted for remains an active
area of theoretical research \citep{Romano07, Decressin09, deMink09,
  Decressin10}.

\section{Expulsion of residual gas and initial cluster radii}

Observations \citep{Lada_Lada03} show that the star-formation efficiency
\begin{equation}
\epsilon = {M_{\rm ecl} \over M_{\rm ecl}+M_{\rm gas}} \approx 0.3. 
\end{equation}
In pure gravo-hydrodynamical simulations of star formation most of the
gas can get accreted onto the proto stars (i.e. sink particles in the
computations), and $\epsilon$ is determined by when the computation is
halted.  The first simulations of star formation in a turbulent
magnetised cloud plus stellar feedback \citep{PB09, Peters11,
  Peters11a}, however, now demonstrate that $\epsilon$ is small on a
global cloud scale, because the magnetic field stabilises the cloud on
large scales ($>1\,$pc) while the feedback inhibits fragmentation on
small scales. Since virtually all clusters older than about one~Myr
are free of gas, this can only mean that the residual gas amounting to
about 70~per cent of the original cloud core mass is expelled from the
cluster within less than~one~Myr.

A simple estimate of the cloud core binding energy, in comparison with
the energy imparted by the feedback energy from massive stars, shows
that within less than a crossing time an order of magnitude more
energy is deposited in the cloud core. The parameters taken here are a
cloud core of radius 1~pc and masses $10^4$ and $10^5\,M_\odot$
leading to binding energies of $8.6\times 10^{48}\,$erg and $8.6\times
10^{50}\,$erg, respectively. The corresponding crossing times are~0.48
and~0.15~Myr. A single pre-supernova star of $15\,M_\odot$ injects in
total $3\times 10^{50}\,$erg while a single $85\,M_\odot$ star injects
$3\times 10^{50}\,$erg into the cloud, within~0.1~Myr
\citep{Maeder90}.  Thus, the disruption of the nebula is most probably
rapid if not explosive, within less than a crossing time scale. Only
in massive clusters with $M_{\rm ecl}>10^5\,M_\odot$ is a more
adiabatic evolution probably the case \citep{TenarioTagle03, BKP08},
because neither individual core-collapse supernovae nor the feedback
of all OB stars in the cluster contain sufficient energy to unbind the
gas, parts of which shock-thermalises and may form further stars. The
effective $\epsilon$ probably therefore increases with increasing
$M_{\rm ecl}>10^5\,M_\odot$.

There is ample observational evidence for explosive gas expulsion. The
ONC is super-virial \citep{KAH,Kroupa05}, while star clusters appear
to show a major expansion in the first~10s~Myr of their life time
(\citealt{Bastian06,GB06,deGP07,Bastian08,Brandner08}, but see also
\citealt{Gieles10}).  Very young embedded clusters, or CSFEs, appear
to be very compact, with radii of less than a~pc, as is inferred from
direct observations \citep{Kroupa05} and inverse dynamical population
synthesis (Marks \& Kroupa 2012, submitted). Inverse dynamical
population synthesis is a potentially powerful tool to infer the
initial conditions of star clusters by assuming the initial binary
population to be invariant. The present-day binary population stores
the dynamical history of the population and the deepest negative
energy state of the cluster can be read off from the energy
distribution of the present day binary population.

Such work leads to the result that there is a weak
embedded-cluster-mass--radius relation for pre-gas-expulsion CSFEs
(Marks \& Kroupa 2012, submitted)
\begin{equation}
r_h = 0.11^{0.08}_{0.05} M_{\rm ecl}^{0.12\pm0.05},
\label{eq:pk_mr}
\end{equation}
where $M_{\rm ecl}$ is the stellar mass and $r_h$ is the half-mass
radius.  It will have to be seen how this result, which is obtained by
inferring the initial density given the present-day energy
distribution of the binary population in each considered cluster, can
be made conform to \cite{Pfalzner09}'s suggestion that there are two
sequences of birth CSFEs following different density--radius scaling
relations.

Concerning Eq.~\ref{eq:pk_mr}, a metallicity dependence also emerges
in that metal-poor clusters appear to have had a systematically
smaller $r_h$ at a given mass $M_{\rm ecl}$ than metal richer clusters
\citep{MK10}.  This may suggest that metal-poor gas can collapse to
denser states while the more efficient coupling of photons to more
metal-enriched gas increases the opacity and thus limits the depth of
the collapse while also increasing fragmentation. This is the same
physical reason for stellar winds being metal dependent and
star-formation theory predicting a bottom heavy stellar IMF for metal
rich star-forming conditions (e.g. \citealt{Bastian10}). This finding
would imply that metal rich CSFEs with $M_{\rm ecl}>10^4\,M_\odot$ are
less likely to form bound GCs in present-day galaxy--galaxy
encounters, as has been surmised by \cite{Forbes97}, because by
forming with larger $r_h$ they are more susceptible to damage from gas
expulsion and tidal perturbations.

Taking gas-expulsion into account, it has been shown that one single
pre-ONC CSFE evolves, via gas expulsion, through the ONC at an age of
about 1~Myr, to the Pleiades at an age of about 100~Myr and then onto
to the Hyades and Praesepe at ages of a few hundred~Myr \citep{KAH,
  Portgies01}.  Even the binary and stellar populations, as well as
the density and velocity profiles come out remarkably well in the
models compared to the real clusters. It is remarkable that all these
clusters at a mass scale of about $1000\,M_\odot$ are so similar. We
will return to this further below (Sec.~\ref{sec:pk_CMF}).

\section{Thickening galactic disks, chain galaxies, stellar streams
  and dark star clusters}

The origin of thick disks of galaxies remains unclear, and the reason
why the MW thin disk thickened with age also remains unresolved.  GAIA
data will allow significant constraints to be placed on the acting
heating mechanisms through a precise quantification of the phase-space
distribution of the various stellar components.  A popular hypothesis
is that dark-matter sub-clumps and merging dark-matter dominated
satellite galaxies thicken the disks. This scenario would be natural
for the popular cold- or warm-dark-matter based cosmological models
because they are synonymous with hierarchical galaxy formation and
imply, by logical necessity, the harassment of thin galactic disks by
myriads of dark matter sub clumps. It has proven to be difficult to
account for the observed heating though, since the theoretically
implied impact of the sub-structures would be far more damaging if not
destructive to the thin disks.  \cite{Sellwood10} discusses the
various heating processes acting within galactic disks such as heating
by bars, transient spiral patterns, molecular clouds, but the full
heating has not been accounted for.

Young star clusters that expel their gas explosively loose a dominant
fraction of their stellar population which expands, to first order,
with the pre-gas-expulsion velocity dispersion. Thus, if the velocity
dispersion in the embedded cluster is
\begin{equation}
\sigma \approx \left( {G\,M_{\rm ecl} \over \epsilon \, R} \right)^{1\over2},
\end{equation}
where $G$ is the gravitational constant, then $\sigma>20\,$km/s if
$M_{\rm ecl}> 10^5\,M_\odot$ for $\epsilon\,R\approx 1\,$pc. These are
realistic quantities since $r<3\,$pc and $\epsilon\approx 0.33$,
suggesting that explosive gas expulsion may lead to thickened
disks. \cite{Kroupa02} shows that the thickening history of the MW
disk can be accounted for by clustered star formation if the
star-formation rate (SFR) decreased over time therewith forming star
cluster populations with maximal cluster masses that decrease with
decreasing SFR \citep{WKL04}. That the thick disk of the MW could have
formed via ``popping'' CSFEs of mass $\approx 10^6\,M_\odot$ has been
shown with simulations by \cite{Assmann11}.

This model for the thickening of the MW disk is naturally consistent
with the observed {\it chain galaxies} which show edge-on disk
galaxies at high redshift undergoing star formation bursts in many
knots within their disks \citep{EE06}.

The relevance for GAIA is that each CSFE must be associated with a
population of stars spreading apart along a thick tidal stream which
resulted from the gas-expulsion event. A re-virialised star cluster
may be the core of this expanding population, and if the CSFE did
yield a bound star cluster then this cluster will be forming a thin
kinematically cold tidal stream composed of stars that evaporate from
the cluster due to two-body relaxation. These cold streams are of much
interest because they may be used to probe how sub-structured the
putative dark matter halo of the MW really is. It has already emerged,
however, that the observed overdensities in known thin streams can be
fully accounted for by {\it epicyclic overdensities}
\citep{Kuepper10}. A star which is nudged out of its cluster through
one of the Lagrange points orbits the MW on a slightly different orbit
to that of the cluster and thus the angular separation between the
star and the cluster oscillates as the star drifts away from its
cluster. Since many stars perform essentially the same motions they
accumulate in regions where stars turn around relative to the cluster.

This remarkable development, achieved through high-precision $N$-body
computations that have become possible with Sverre Aarseth's Nbody
codes and the GPU-based computing platforms \citep{Aarseth08}, opens
entirely new and extremely powerful tools for constraining the MW
potential and gravitational dynamics in general, since the position
and properties of the epicyclic overdensities are a measure of the
cluster--MW combined effective potential. It is expected that the GAIA
data will contain many thin streams, given that the entire MW field
population of stars stems from dissolving CSFEs. It will become
possible to measure how the streams diffuse apart with the age of
their trace population.

Thus, around each star cluster, such as the Pleiades, the GAIA data
ought to reveal two associated kinematical populations. The expanding
population and the evaporated population. The total of the expanding
population is likely to be about 2/3rds of the initial CSFE membership
for the Pleiades \citep{KAH}, for example. 

Furthermore, in each re-virialised post-gas-expulsion star cluster
there is a competition of time scales: the remnants of massive stars
(neutron stars and black holes) accumulate near the cluster center due
to mass segregation. There they eject each other through three-body
encounters on a core ejection time-scale. At the same time, low-mass
stars evaporate from the cluster on a two-body relaxation time-scale,
which is enhanced in stronger tidal fields. It turns out \citep{BK11}
that the evaporation time scale is faster than the core-ejection
time-scale in star clusters within about 5~kpc distance from the
Galactic center. There, clusters with sufficient mass to contain many
stellar remnants evolve to black-hole dominated objects with a few
stars being bound, the {\it dark star clusters}. GAIA should find such
super-virial (``dark'') clusters if they exist. Therewith the typical
kick a remnant receives during the supernova explosion will be
constrained to a small value as otherwise the cluster looses the
remnants at explosion time. Once the existence of dark clusters has
been observationally proven, it will be ascertained that star clusters
must be the sources of gravitational wave emission \citep{BBK10}.

Mining the GAIA data will therefore open entirely new doors into
understanding the formation and disruption of CSFEs as well as into
gravitational dynamics and the physics of stellar explosions.

\section{The mass function of CSFEs, of globular clusters and the
  population II halo} 
\label{sec:pk_CMF}

Embedded star clusters have been found to have a power-law mass
function (ECMF) with power-law index $\beta\approx 2$ (the Salpeter
index would be~2.35; \citealt{Lada_Lada03}). If residual gas expulsion
is relevant for early cluster evolution, then it follows that the
relation between the final mass of the re-virialised young cluster is
a fraction of the pre-gas expulsion stellar mass,
\begin{equation}
M_{\rm rev} = f_{\rm st} \, M_{\rm ecl},
\end{equation}
where $f_{\rm st}<1$ is the fraction of stars remaining in the
re-virialised cluster. Given that $f_{\rm st}$ is likely to be a
function of the depth of the potential well, 
\begin{equation}
f_{\rm st}={\rm fn}(M_{\rm ecl}),
\end{equation} 
and also of the number of massive stars in the CSFE (being $\propto
M_{\rm ecl}$ for an invariant IMF), it follows that $f_{\rm st}$ may
have a significant minimum around about $M_{\rm
  ecl}=10^4\,M_\odot$. Here the potential well is still sufficiently
shallow while O~stars are already present in the population. It is at
this mass scale that the expulsion of residual gas may be most
damaging and $f_{\rm st}$ may reach a minimum. Such ideas lead to the
result that the post-gas expulsion MF of re-virialised young clusters
have a structured MF which reflects the variation of $f_{\rm st}$ with
$M_{\rm ecl}$.

\cite{KB02} show that this ansatz naturally accounts for the turnover
of the MF of clusters near $10^5\,M_\odot$ as is observed for the
ancient GCs. The gas expulsion process unbinds a large fraction of the
embedded cluster population such that the MW halo population~II
spheroid emerges naturally -- it is composed of the quickly dissolved
low-mass star clusters that formed together with the present-day GCs
plus the stars that were lost from the present day GCs due to residual
gas expulsion.  Such concepts have been incorporated into $N$-body
modeling of cluster populations by \cite{BKP08} and
\cite{Parmentier08}.

The post-gas expulsion cluster MF may also have a pronounced peak near
$10^3\,M_\odot$ where the ONC, Pleiades, Hyades and Praesepe lie. This
may explain why such open clusters may perhaps be overrepresented.

\section{Top-heavy IMF in star bursting clusters}

As discussed above, GAIA data will constrain the gas-expulsion process
from CSFEs. Constraining the gas-expulsion process is of much
importance also for the issue of globular cluster (GC) formation and
the possible variation of the stellar IMF with physical cloud
conditions. The data will thus touch upon a holy grail of star
formation research, namely the finding of conclusive evidence of the
long-expected variation of the IMF with star-forming cloud metallicity
and cloud density.

The deep observations by \cite{deMarchi2007} of 20~GCs show that the
present-day stellar MF (PDMF) becomes increasingly shallower over the
stellar mass range $0.3-0.8\,M_\odot$ with decreasing cluster
concentration. The same trend holds also with increasing
metallicity. Both do not follow readily from theory \citep{MK10}.

That low-concentration clusters are depleted of low mass stars was
surprising because the expectation was the opposite,
i.e. low-concentration clusters ought to retain their low-mass stars
and high concentration (post core collapse) clusters ought to be more
evolved dynamically having lost a larger fraction of their low-mass
stars. Also, the theoretical expectation has been that low-metallicity
clusters ought to have top-heavy IMFs, i.e. should have flatter PDMFs,
contrary to the observations.

If clusters were to form mass segregated but with an invariant
canonical stellar IMF for $<1\,M_\odot$ and filling their tidal radii
then low-mass stars evaporate preferentially immediately and the
observed trend between PDMF and concentration can be reproduced
\citep{BdeMK08}. This comes about because by being initially mass
segregated, initially tidally-filling clusters loose the most weakly
bound (and thus low-mass) stars preferentially.  But this ansatz
cannot reproduce the trend of the PDMF and of the concentration with
metallicity. Also, it is not clear why star clusters should form mass
segregated filling their tidal radii, given that the tidal radii of
the young clusters are very large (e.g. about 120~pc for a
$10^5\,M_\odot$ cluster at a distance of 8~kpc from the MW).

The metallicity trend however gives a clue: If gas expulsion is more
efficient for more metal-enriched gas then CSFEs forming out of such
material would suffer more damage than low-metallicity CSFEs (compare
to metallicity-dependent stellar winds). Indeed, it turns out that
low-concentration clusters with flatter PDMFs result naturally if GCs
were born mass segregated, had radii of about a~pc and expelled their
gas at a rate in dependence of the metallicity \citep{MKB08,MK10}.
The first ever $N$-body computation over the entire life-history of a
low-mass globular cluster, Pal~14, verifies that a viable
re-virialised post-gas-expulsion solution to the low-concentration
cluster Pal~14 suggests it may have had a stellar mass function depleted
in low mass stars and a large half-mass radius of about 20~pc
\citep{Zonoozi11}.

The constraints arrived at for the 20~GCs with deep observations imply
that in order to actually remove the residual gas a certain amount of
feedback needs to be invoked. This feedback is only possible if the
IMF was increasingly top heavy with increasing cluster$+$gas cloud
core density and decreasing metallicity (Marks et al. 2012, submitted).
This may perhaps be the first ever observationally derived evidence
for a density and metallicity dependent IMF variation in consistency
with star-formation theory. The onset of this suggested systematic IMF
variation is at a pre-cluster cloud-core density
$>10^5\,M_\odot/$pc$^3$ and a mass-scale of a CSFE on a~pc scale of
$>10^5\,M_\odot$.

\section{Conclusions}

An outline of some of the work done on the physical processes that
shape young star clusters and which are relevant for GAIA by leaving
imprints in the kinematical field of the MW has been given. An
analysis of GAIA astrometric data across the MW is likely to
unambiguously constrain the gas expulsion process from CSFEs, because
it leaves characteristic signatures in the MW disk. Once the gas
expulsion process has been constrained, it will emerge whether
galactic disks can be thickened by residual gas expulsion from compact
CSFEs and whether the star-cluster MF undergoes the rapid
transformation from a power-law embedded cluster MF to a structured MF
for older re-virialised clusters. Also, it will then become evident
whether the systematic IMF variation, deduced from the careful
dynamical analysis of deep GC observations, is supported, since the
analysis rests on the notion that residual gas expulsion is a major
physical process governing the emergence of star clusters from their
embedded phase. It need not be overemphasised that verification of a
systematically variable IMF has profound cosmological implications.

The GAIA mission will thus provide fundamental constraints not only on
gravitational dynamics and the potential of the MW, but will also
constrain the star-formation process on a pc~scale and how it affects
the morphology of galaxies.

%
%
\small  
%
\section*{Acknowledgments}   
I would like to sincerely thank the organisers for a most memorable
conference, and Michael Marks for proof reading this manuscript.  
%
%
%
%

\bibliographystyle{aa}
\bibliography{mnemonic,Kroupa_P_ref}

\end{document}